\documentclass{svproc}
\usepackage{url}
\usepackage{algpseudocode}
\usepackage{amsfonts} 
\usepackage{graphicx}

\begin{document}
\renewcommand\thefootnote{}% Suppresses numbering for the footnote
\mainmatter              % start of a contribution
\title{A topology-based algorithm for the isomorphism check of 2-level Orthogonal Arrays}
\titlerunning{A Topology-based algorithm for isomorphism check}  % abbreviated title (for running head)
%                                     also used for the TOC unless
%                                     \toctitle is used
%
\author{Roberto Fontana \inst{1} \and Marco Guerra \inst{2}}
\authorrunning{R. Fontana, M. Guerra} % abbreviated author list (for running head)
%
%%%% list of authors for the TOC (use if author list has to be modified)
%\tocauthor{Roberto Fontana \and Marco Guerra}
%
\institute{Department of Mathematical Sciences, Politecnico di Torino \\ Corso Duca degli Abruzzi 24, 10129, Torino, Italy\\
\email{roberto.fontana@polito.it},
\and
Institut Fourier, Universit\'{e} Grenoble-Alpes, CNRS \\
100, Rue des Mathematiques, 38610 Gières \\
\email{marco.guerra@univ-grenoble-alpes.fr}
}

\maketitle              % typeset the title of the contribution

\begin{abstract}
We introduce a construction and an algorithm, both based on Topological Data Analysis (TDA), to tackle the problem of the isomorphism check of Orthogonal Arrays (OAs). Specifically, we associate to any binary OA a persistence diagram, one of the main tools in TDA, and explore how the Wasserstein distance between persistence diagrams can be used to inform whether two designs are isomorphic. 
% We would like to encourage you to list your keywords within
% the abstract section using the \keywords{...} command.
\end{abstract}
\section{Introduction}
\footnotetext{This preprint has not undergone peer review or any post-submission improvements or corrections. The Version of Record of this contribution has been accepted for publication in Springer, Methodological and Applied Statistics and Demography I. SIS 2024, Short Papers, Plenary and Specialized Sessions - Ed. A. Pollice, P. Mariani (to appear Nov 24)}
Factorial designs are commonly used in various fields of application, including engineering and agriculture. The isomorphism check determines whether two designs are equivalent. For qualitative factors, two designs are combinatorially isomorphic if one design can be obtained from the other by (i) reordering the runs, (ii) relabeling the factors and/or (iii) switching the levels of one or more factors. For quantitative factors, two designs are geometrically isomorphic if one design can be obtained from the other by (i), (ii), and/or reversing the level order of one or more factors. For 2-level designs, combinatorial and geometric isomorphism are equivalent. The isomorphism check has been extensively studied in the statistical literature (see \cite{clark2001equivalence,lin2012isomorphism,angelopoulos2007effective,pang2011geometric,lin2008isomorphism,wu2001generalized}), and is a known hard problem computationally.

In this work we focus on the isomorphism check for \emph{binary} factorial designs, specifically, we work on the subclass of Orthogonal Arrays of strength 2. However, our method can be applied to OAs of any strength or, more generally, to any type of 2-level designs. We consider OAs because the set of non-isomorphic OAs for different number of factors is easily available (e.g. we used the Orthogonal Array package, \cite{schoen2010complete}), so the performance of our algorithm could be easily assessed. 

We introduce a simple construction that associates to any $d$-variate Bernoulli random variable a \textit{persistence diagram}, a key tool in TDA. By doing so, and by interpreting an OA as a multivariate Bernoulli, we make use of a notion of Wasserstein distance between persistence diagrams to inform the isomorphism check problem, via a statistical procedure. Notice the 2-level hypothesis is required for the construction to work. We report good performance of our algorithm in low-dimensional cases.

\section{The isomorphism check problem}
\subsection{Factorial designs and Orthogonal Arrays}
Given $d$ binary factors $A_1,\ldots,A_d$, whose levels are coded as $\{0,1\}$, we represent a fraction $\mathcal{F}$ of the full factorial design $\mathcal{D}=\{\alpha \in \{0,1\}^d\}$ using the counting vector $v_{\mathcal{F}}$.  Each component $v_\alpha$ of the counting vector $v_{\mathcal{F}}$ is the number of times that the point $\alpha$ appears in $\mathcal{F}$, $v_{\mathcal{F}}=(v_\alpha,\alpha \in \{0,1\}^d)$.  It follows that $v_\alpha \geq 0,\alpha \in \{0,1\}^d$. We order the set $\{\alpha \in \{0,1\}^d\}$ lexicographically.

\textbf{Pmf and moments} We observe that each vector $v_{\mathcal{F}}$ determines $p_\mathcal{F}$, the probability mass function (pmf)  of a $d$-variate Bernoulli random variable $X=(X_1,\ldots,X_d)$ simply by dividing each component of $v_{\mathcal{F}}$  by $N$, the number of points contained in $\mathcal{F}$: 
$p_\mathcal{F}=\frac{1}{N} v_{\mathcal{F}}$.
We also consider $\mu_\mathcal{F}$, the vector of moments of $X \sim p_{\mathcal{F}}$, $\mu_\mathcal{F}=(\mu_\alpha, \alpha \in \{0,1\}^d)$, where $\mu_\alpha=E[X^\alpha] \equiv E[X_1^{\alpha_1} \cdot \ldots \cdot X_d^{\alpha_d}]$.  

%For example for $d=2$ the counting vector $v_{\mathcal{F}}$ of the fraction $\mathcal{F}=\{(0,0),(1,0),(0,0)\}$ is $v_{\mathcal{F}}=(2,0,1,0)$, the corresponding pmf is $p_{\mathcal{F}}=(2/3,0,1/3,0)$ and $\mu=(1,1/3,0,0)$.
For example, for $d=2$, the counting vector $v_{\mathcal{F}}$ of $\mathcal{F}=\{(0,0),(0,0),(0,1),$ $(1,0),(1,1),(1,1)\}$ is $v_{\mathcal{F}}=(2,1,1,2)$,  the corresponding pmf is $p_{\mathcal{F}}=(1/3,1/6,$ $1/6,1/3)$, and $\mu_{\mathcal{F}}=(1,1/2,1/2,1/3)$.

\textbf{OAs} An OA of strength $t$ on $N$ runs, with $d$ binary factors, $1 \leq t \leq d$, is a fraction $\mathcal{F}$ of $N$ possibly repeating points of $\mathcal{D}=\{0,1\}^d$ such that
every projection $\mathcal{F}_t$ of $\mathcal{F}$ on a subset of $t$ factors is a full factorial design with $t$ factors and each run of $\mathcal{F}_t$ appears the same number of times, $N/2^t$.
The fraction $\mathcal{F}$ above is an OA of strength $t=1$ and size $N=6$.

\textbf{Isomorphism} Two fractions $\mathcal{F}_1$ and $\mathcal{F}_2$ of $\mathcal{D}$ are said to be isomorphic if one is obtained from the other by a combination of (i) reordering the runs, (ii) relabeling any of the factors, (iii) switching the levels on any factor.

For example the OA  $\mathcal{F}_1=\{(0,1),(0,1),(0,0),(1,1),(1,0),(1,0)\}$ is isomorphic to $\mathcal{F}$ above; it has been obtained from $\mathcal{F}$ by switching the levels on factor $A_2$.

\subsection{Topological Data Analysis}
Topological data analysis (TDA) is a recent subject of mathematics that leverages classical tools from algebraic topology, such as homology theory, to study data. In the last 25 years it has found wide application, partly thanks to its sound theoretical grounding (see \cite{chazalIntroTDA} for a review). The main tool in TDA is called persistent homology (PH); intuitively, homology describes the \textit{shape} of a topological object via algebraic invariants. Data, in a very broad sense, can be given the structure of a family of topological spaces, each described by homology. Persistent homology, then, describes the \textit{shape of data} by tracking the evolution of homology through this family of spaces associated with data. \\
A proper introduction of PH cannot fit here, therefore we refer to the bibliography (\cite{EdelsbrunnerIntro}) and only sketch the concepts we will use. PH uses the homology of simplicial complexes, which are discrete topological spaces based on standard building blocks called simplices, suitable for computation. Data in a metric space yields an \textit{increasing family} of simplicial complexes, called a \textit{filtration}, i.e. a family of simplicial complexes that are totally ordered by inclusion. Simplicial homology counts the \textit{obstructions to connectivity}, that is, connected components in dimension 0, holes in dimension 1, voids in dimension 2, and so on. Tracking simplicial homology along a filtration yields persistent homology, that is the birth and death of connected components, holes, voids and so on. This information is summarized in the \textit{persistence diagram}, which we denote by $Dgm$, a set of points in the plane, each referring to a homological feature (hole, void, etc), whose first coordinate is its birth time and second coordinate its death time. Note features that persist until the end of the filtration have infinite second coordinate. \\
\subsection{The Dgm of an OA}
The full factorial design $\mathcal{D}=\{\alpha \in \{0,1\}^d\}$ is a simplicial complex, namely the $d-1$ simplex plus all its faces. Intuitively, introduce $d$ abstract vertices: each $\alpha$ denotes which vertices are "switched on" (have a 1), and therefore defines a simplex, in particular a face of the $d-1$ simplex. Given a Bernoulli pmf on $\mathcal{D}$, the moments vector $\mu$ associates a real number $\mu_\alpha$ to each simplex $\alpha$. We state here without proof that the sublevel sets of the function $1 -\mu_\alpha : \mathcal{D} \rightarrow \mathbb{R}$ define a filtration of simplicial complexes. Therefore, to each pmf we can associate a well-defined persistence diagram. By the above, we can do the same to any binary OA.

\begin{figure}
    \centering
    \includegraphics[height=4.5cm, width=0.55\textwidth]{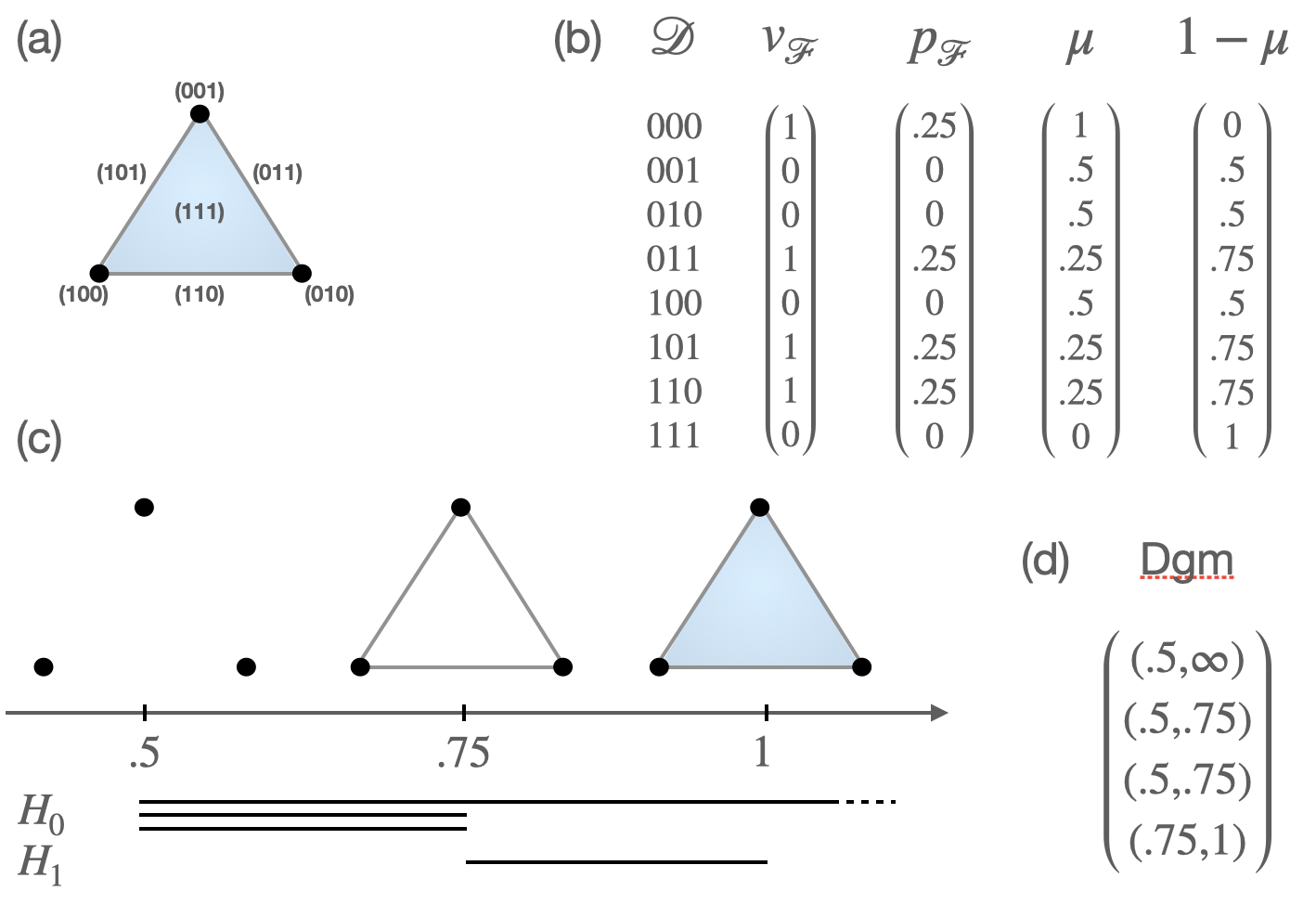}
    \caption{(a) The simplices corresponding to the $\alpha \in \{0,1\}^3$. (b) A 4-point OA in $d=3$, and its corresponding pmf, moment vector, and filtration $1-\mu$. Notice the $\alpha$'s are ordered lexicographically in the vectors. (c) The corresponding filtration made of sublevel sets of $1-\mu$. The bars represent persistence of homological features (3 connected components and one loop). (d) The persistence diagram showing the pairs.}
    \label{fig:corresp}
\end{figure}

%\FloatBarrier
\subsection{Wasserstein distances}
The Wasserstein distance is a distance between two probability measures  over a common metric space. In this work we consider the Wasserstein distance  $d^1_W(p_1,p_2)$ between two pmfs, $p_1$ and $p_2$, that correspond to the fractions $\mathcal{F}_1$ and $\mathcal{F}_2$, respectively. The distance is computed by finding an optimal permutation $\pi$, that maps the two vectors as close as possible
\[
 d^1_W(p_1,p_2) := \displaystyle \inf_{\pi} \sum_{\alpha \in \{0,1\}^d}  \vert p_1(\alpha) - p_2(\pi(\alpha)) \vert .
\]
If (and only if) $d^1_W(p_1,p_2) = 0$, then the two pmfs are a permutation of one another. It follows that if $d^1_W(p_1,p_2) >0$ then $\mathcal{F}_1$ and $\mathcal{F}_2$ are not isomorphic.

The Wasserstein distance between two persistence diagrams $Dgm1$ and $Dgm_2$ corresponding to $\mathcal{F}_1$ and  $\mathcal{F}_2$, respectively, is obtained by finding a bijection between the finite points of the diagrams as points in $\mathbb{R}^2$, where the metric is the Euclidean one.
\[
    d^2_W(Dgm_1, Dgm_2) := \displaystyle \inf_{\eta : Dgm_1 \leftrightarrow Dgm_2} \left( \sum_{x \in Dgm_1} \Vert x - \eta(x) \Vert_2 \right)
\]

   If (and only if) $d^2_W(Dgm_1, Dgm_2)=0$, then the two PD's are a permutation of one another. This implies the filtration vectors are a \textit{pairwise} permutation of one another, i.e. a permutation that is compatible with the pairing of simplices given by the persistence pairs. 

\section{The algorithm}
We propose the following statistical algorithm to test if two OAs, $\mathcal{F}_1$ and  $\mathcal{F}_2$ are isomorphic. We denote by $p_i$ ($Dgm_i$)  the pmf (the persistence diagram) corresponding to $\mathcal{F}_i$, $i=1,2$, respectively, and by $T$ the maximum number of iterations. 

Given  $\mathcal{F}_1$ and  $\mathcal{F}_2$ the corresponding pmf's $p_1$ and $p_2$ are computed. If the Wasserstein distance $d^1_W(p_1,p_2)$ between $p_1$ and $p_2$ is positive a permutation $\pi$ such that $p_2=\pi(p_1)$ does not exist and then $\mathcal{F}_1$ and  $\mathcal{F}_2$ are not isomorphic. If $d^1_W(p_1,p_2)=0$ a permutation $\pi$  such that $p_2=\pi(p_1)$ exists. We must decide if the permutation is a combination of (i) reordering the runs, (ii) relabeling any of the factors, (iii) switching the levels on any factor or not. To do that we compute $d^2_W(Dgm_1, Dgm_2)$, the Wasserstein distance between the persistence diagrams $Dgm1$ and $Dgm_2$ corresponding to $\mathcal{F}_1$ and $\mathcal{F}_2$, respectively. If the distance is zero the algorithm indicates that  $\mathcal{F}_1$ and  $\mathcal{F}_2$ are isomorphic. If the distance is greater than zero the algorithm considers $T$ fractions $\mathcal{F}_2^{(1)} , \mathcal{F}_2^{(2)}, \ldots, \mathcal{F}_2^{(T)}$ isomorphic to $\mathcal{F}_2$ and the corresponding persistence diagrams $Dgm_2^{(1)}, Dgm_2^{(2)}, \ldots, Dgm_2^{(T)}$. If any of the distances  $d^2_W(Dgm_1, Dgm_2^{(t)}), t=1,\ldots,T$  is zero than the algorithm indicates that $\mathcal{F}_1$ and  $\mathcal{F}_2$ are isomorphic, otherwise $\mathcal{F}_1$ and  $\mathcal{F}_2$ are considered as non-isomorphic.

\begin{figure}
    \centering
\begin{algorithmic}
\footnotesize
\State  Input $p_1 , p_2$, and $T$
\State  Output: True/False, where True means that $\mathcal{F}_1$ and  $\mathcal{F}_2$ are isomorphic.
\If {$d^1_W(p_1, p_2) > 0$} 
	\State Status:=False; 
\Else
	\State Status:=False
	\For {$i = 1, \dots, T$}
		\State Compute $f_1, f_2$
		\If {$d^1_W(f_1, f_2) == 0$} 
			\State Compute $Dgm_1, Dgm_2$
			\If {$d^2_W(Dgm_1, Dgm_2) == 0$}
				\State  Status:=True
 			\EndIf
			\State $p_2 \leftarrow$ Sample $p_2'$ isomorphic to $p_2$ 
		\EndIf
	\EndFor	
\EndIf
\State return Status; 
\end{algorithmic}
\label{alg:algo1}
\end{figure}

\section{Applications}
\begin{figure} 
\centering
    \includegraphics[width=0.45\textwidth]{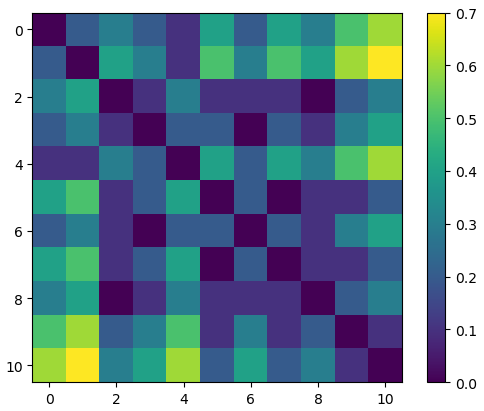}
    \includegraphics[width=0.4\textwidth, height=4.8cm]{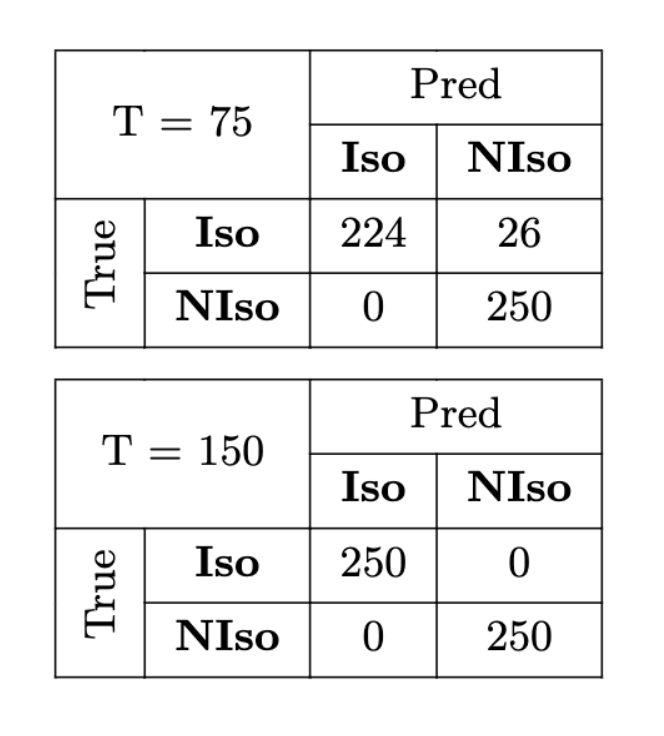}
    \caption{Left, $d^1_W$ between pmfs of the 11 classes for $d=5$, $N=20$, $t=2$. Right, the classification results of the algorithm for a sample of 500 OAs of the same parameters for $T=75$ and $T=150$. The second case achieves perfect classification. We sample from a set of $10,752$ OAs.}
\label{fig:d1}
\end{figure}

We consider binary OAs for $d=5$, size $N=20$, and strength $t=2$.
There are 11 classes of non-isomorphic OAs for these parameters. We denote by $\mathcal{F}_0 \ldots, \mathcal{F}_{10}$ the 11 non-isomorphic OAs that we have computed using the  Orthogonal Array package, \cite{schoen2010complete} and by $p_0 \ldots, p_{10}$ the corresponding pmfs. We observe (see Figure \ref{fig:d1} left) that there are cases where the Wasserstein distance between pmfs fails to discriminate non-isomorphic arrays; for example the $d^1_W(p_3,p_6)=0$. Conversely, in all cases the criterion based on the $d^2_W$ between persistence diagrams is able to correctly classify isomorphic/non-isomorphic OAs (see Figure \ref{fig:d1} right). We have also tested the algorithm in the cases  $d=3,4$ ($N=20$, $t=2$), with equally correct results.

\section{Concluding remarks}
To the best of our knowledge, this is the first time a TDA-based methods has been applied to solving the isomorphism check for binary orthogonal arrays. For sufficiently large $T$, the algorithm was successful in all cases for $d=3,4,5$, $N=20$, $t=2$ that we have considered. Our main goal is to prove that the statistical procedure is in fact exact at least in one direction:
\begin{proposition}[Conjecture]
    If $\mathcal{F}_1$ and $\mathcal{F}_2$ are not isomorphic, then  
\[
d^2_W(Dgm_1, Dgm_2) > 0
\]
\end{proposition}

%%
%% ---- Bibliography ----
%%

%\bibliographystyle{..\\styles\\bibtex\\spmpsci}
\bibliographystyle{spmpsci}
\bibliography{biblioRFMG}

\end{document}